\documentclass[a4paper,aps,preprint,nofootinbib]{revtex4}
\pdfoutput=1
\usepackage[latin1]{inputenc}
\usepackage{bbm}
\usepackage{bm}
\usepackage{graphicx}
\usepackage{epsfig}
\usepackage{subfigure}
\usepackage{latexsym}
\usepackage{amsmath}
\usepackage{amsfonts}
\usepackage{amssymb}
\usepackage{wasysym}
\usepackage{color}

\newcommand{\be}{\begin{equation}}
\newcommand{\ee}{\end{equation}}

\newcommand{\bea}{\begin{eqnarray}}
\newcommand{\eea}{\end{eqnarray}}

\definecolor{rossoCP3}{cmyk}{0,.88,.77,.40}
\begin{document}


\title{LHC Data and Aspects of New Physics}
\author{Tommi Alanne $ ^{a,b}$\footnote{tommi.alanne@jyu.fi}}
\author{Stefano Di Chiara$ ^{a}$\footnote{stefano.dichiara@helsinki.fi}}
\author{Kimmo Tuominen $ ^{a,b}$\footnote{kimmo.i.tuominen@jyu.fi}}
\affiliation{\mbox{ $ ^{a}$ Helsinki Institute of Physics, P.O.Box 64, FI-000140, Univ. of Helsinki, Finland}} 
\affiliation{\mbox{ $ ^{b}$ Department of Physics, P.O.Box 35, FI-40014, Univ. of Jyv\"askyl\"a, Finland} }

\begin{abstract}
We consider the implications of current LHC data on new physics with strongly interacting sector(s). We parametrize the relevant interaction Lagrangian and study the best fit values in light of current data. These are then considered within a simple framework of bosonic technicolor. We consider first the effective Lagrangian containing only spin-0 composites of the underlying theory, which corresponds to a two Higgs doublet model. With respect to this baseline, the effects of the vector bosons, a staple in strong interacting theories, are illustrated by considering two cases: first, the case where the effects of the vector bosons arise only through their mixing with the electroweak SU(2)$_L$ gauge fields and, second, the case where also a direct interaction term with neutral scalars exists. We find that the case of a $W'$ coupling to the Higgs boson only via the mixing of vector fields produces a negligible improvement in the fit of the present data, while even a small direct coupling of the composite vector fields to the Higgs allows the tested model to fit optimally the experimental results. 
\end{abstract}
\maketitle

\section{Introduction}
\label{Intro}
The recent discovery of a light scalar with properties compatible with those of the Standard Model (SM) Higgs boson, $h^0$, imposes new experimental tests on previously viable beyond the Standard Model (BSM) theory frameworks. Fervent activity in this direction has focused mostly on the possibility that a new charged particle could enhance the Higgs decay rate to two photons \cite{Azatov:2012bz,Carmi:2012yp,Espinosa:2012ir,Giardino:2012ww,Giardino:2012dp,Carmi:2012in,Carena:2012xa,Ellis:2012hz,ArkaniHamed:2012kq, Bonnet:2012nm, Banerjee:2012xc, Corbett:2012dm, Corbett:2012ja}. This is measured at LHC and Tevatron, with the full dataset, to be slightly enhanced compared to the SM prediction \cite{ATLASgamma,CMSgamma,:2012zzl}, with the diphoton signal strength equal to $1.65\pm 0.32$ at ATLAS and  $1.11\pm 0.31$ at CMS.\footnote{We use the more traditional "cut based" CMS result for the Higgs decay to diphoton.} While most of the efforts were focused on the Higgs physics associated with a new charged scalar or a vector fermion, both of which naturally arise in supersymmetry \cite{Carena:2012gp,Arbey:2012dq,Cao:2012fz, Christensen:2012ei,Gunion:2012zd,King:2012is,Ellis:2012aa} or composite Higgs frameworks \cite{Contino:2011np,Espinosa:2012qj,Haba:2012zt,Frandsen:2012rj}, less extensive research has been conducted recently on the possibility that a heavy charged vector boson be responsible for the observed deviations of the Higgs couplings from the corresponding SM predictions \cite{Carmi:2012in,Du:2012vh,Abe:2012fb, Andersen:2011nk}. 

A heavy charged vector boson is naturally predicted by phenomenological theories featuring additional gauge groups, like Technicolor, Little Higgs, and Kaluza-Klein models \cite{Dimopoulos:1979es,Farhi:1980xs,ArkaniHamed:2002qy,ArkaniHamed:2002qx,Randall:1999ee,ArkaniHamed:1998rs}. In this paper we want to explore the effect of couplings of a heavy $W'$ boson to the Higgs on LHC and Tevatron observables. To approach this task we first, in Section \ref{datafit}, perform a fit with the parameters of a simple effective Lagrangian featuring rescaled Higgs couplings to SM particles as well as to a heavy charged scalar and a heavy charged vector boson. To have a concrete model to test against the results of the general fit, in Section \ref{2HDMfit} we introduce a low energy effective Lagrangian for a simple bosonic technicolor model, the bosonic Next to Minimal Walking Technicolor (bNMWT) \cite{Sannino:2004qp,Dietrich:2005jn,Belyaev:2008yj,Antola:2009wq}. 
The low energy effective Lagrangian for the scalar sector of bNMWT corresponds to a Type-I Two Higgs Doublet Model (2HDM) \cite{Branco:2011iw}. We scan the allowed parameter space of this model for data points viable under direct search constraints and electroweak (EW) precision tests, compare the data points to the measured Higgs physics observables, and find the optimal fit of the model. In Section \ref{Wpfit} we introduce two composite vector boson triplets in the low energy Lagrangian, while conserving gauge invariance at the microscopic level, which both mix with the SM $W^\pm$ and directly couple to $h^0$, and repeat the goodness of fit analysis above to determine the optimal values of the $W'$ and $W''$ couplings to $h^0$.
We illustrate separately the features which originate from the mixing and from the direct coupling. Our essential conclusion is that mixing alone produces a negligible improvement of the fit to the current data. A direct interaction on the other hand makes the bNMWT model a perfectly viable candidate. 

\section{LHC and Tevatron Data Fit}
\label{datafit}

The experimental results are expressed in terms of the signal strengths, defined as
\be
\hat{\mu}_{ij}=\frac{\sigma_{\textrm{tot}}{\textrm{Br}}_{ij}}{\sigma_{\textrm{tot}}^{\textrm{SM}} \textrm{Br} ^{\textrm{SM}}_{ij}}\ ,\quad \sigma_{\rm tot}=\sum_{\Omega=h,qqh,\ldots}\epsilon_\Omega\sigma_{\omega\rightarrow \Omega}\ ,\quad \omega=pp,p\bar{p} \ ,
\label{LHCb}\ee
where $\epsilon_\Omega$ is the efficiency associated with the given final state $\Omega$ in an exclusive search, while for inclusive searches one simply has $\sigma_{\rm tot}=\sigma_{pp\rightarrow h(X)}$, the $h$ production total cross section.
The signal strengths from ATLAS, CMS, and Tevatron are given in Table~\ref{datatable}.
\begin{table}[htb]
\begin{tabular}{|c||c|c|c|}
\hline
$ij$ & ATLAS & CMS & Tevatron \\
\hline
$ZZ$ & $\,1.50\pm 0.40\,$ & $\,0.91\pm 0.27\,$ &  \\
 $\gamma\gamma$ & $1.65\pm 0.32$ & $1.11\pm 0.31$ & $6.20\pm 3.30$ \\
$WW$ & $1.01\pm 0.31$ & $0.76\pm 0.21$ & $0.89\pm 0.89$ \\ 
$\tau\tau$ & $0.70 \pm 0.70$ & $1.10\pm 0.40$ & \\
$bb$ & $-0.40\pm 1.10$ & $1.30\pm 0.70$ & $1.54\pm 0.77$\\
\hline
\end{tabular}
\caption{Data on inclusive channels from LHC and Tevatron experiments.}
\label{datatable}
\end{table}
All results for Higgs decays to bosons at ATLAS \cite{ATLASgamma,ATLASW,ATLASZ} and CMS \cite{CMSgamma,CMSW,CMSZ}, as well as the decay to $\tau\tau$ at CMS  \cite{CMStau} and the Tevatron results \cite{:2012zzl}, use the full respective dataset, while results for the decay to $b\bar{b}$ at ATLAS  \cite{ATLASb} and CMS \cite{CMSb}, and to $\tau\tau$ at ATLAS  \cite{ATLAStau}, use $12-13\ \rm fb^{-1}$ of integrated luminosity at 8 TeV and $5\ \rm fb^{-1}$ at 7 TeV. The $b\bar{b}$ quark pair is produced in association with a vector boson, with an efficiency assumed equal to one, while the other searches are inclusive. We also include in our analysis the dijet associated $\gamma\gamma$ production based on the full dataset \cite{ATLASgj,CMSgj}, with  signal strengths and efficiencies\footnote{We chose to include only the loose categories from the ATLAS and CMS dataset at 8 TeV.} listed in Table~\ref{ggjjdata}.

\begin{table}[htb]
\begin{tabular}{|c||c|c|c|c|}
\hline
 & ATLAS 7TeV & ATLAS 8TeV & CMS 7TeV & CMS 8TeV \\
\hline
$\gamma\gamma J J$ & $\,2.7\pm 1.9\, $ & $2.8\pm 1.6$ &  $2.9\pm 1.9$ &  $0.3\pm 1.3$ \\
\hline
 $pp\rightarrow h$ & $22.5\%$ & $45.0\%$ & $26.8\%$  & $46.8\%$ \\
$pp\rightarrow qqh$ & $76.7\%$ & $54.1\%$ & $72.5\%$  & $51.1\%$ \\ 
$pp\rightarrow t\bar{t}h $ & $0.6\%$ & $0.8\%$ & $0.6\%$ & $1.7\%$ \\
$pp\rightarrow Vh $ & $0.1\%$ & $0.1\%$ & $0\%$ & $0.5\%$ \\
\hline
\end{tabular}
\caption{Data on exclusive channels from LHC experiments.}
\label{ggjjdata}
\end{table}

Within the class of models we will study, the part of the Lagrangian relevant for the recent LHC data is of the form
\bea
{\cal{L}}_{\textrm{eff}} &=& a_V\frac{2m_W^2}{v_w}hW^+_\mu W^{-\mu}+a_V\frac{m_Z^2}{v_w}hZ_\mu Z^\mu
-a_f\sum_{\psi=t,b,\tau}\frac{m_\psi}{v_w}h\bar{\psi}\psi\nonumber \\
&&+a_{V'}\frac{2m^2_{W'}}{v_w}hW^{\prime +}_\mu W^{\prime -\mu}-a_S\frac{2m_S^2}{v_w}hS^+ S^-,
\label{efflagr}
\eea
where the third and fourth terms involve, respectively, charged (but color singlet) vector and scalar bosons. We fix the mass parameters to the physical mass of the corresponding particle and $v_w$ to the EW vacuum expectation value (vev), $v_w=246$ GeV.

Consequently, the cross sections and branching rates relevant for Higgs physics are related to the corresponding quantities of the SM in a simple way. We define
\be
\hat{\Gamma}_{ij}\equiv \frac{\Gamma_{h\rightarrow ij}}{\Gamma^{\rm SM}_{h_{\textrm{SM}}\rightarrow ij}}\ ,\quad \hat{\sigma}_\Omega\equiv \frac{\sigma_{\omega\rightarrow \Omega}}{\sigma_{\omega\rightarrow \Omega}^{\rm SM}},
\label{hatgs}\ee
and then, in terms of the coupling coefficients in Eq.~(\ref{efflagr}), we have
\bea
\hat{\sigma}_{hqq}=\hat{\sigma}_{hA}=\hat{\Gamma}_{AA}=|a_V|^2\ &,&\quad \hat{\sigma}_{h\bar{t}t}=\hat{\sigma}_{h}=\hat{\Gamma}_{gg}=\hat{\Gamma}_{\psi\psi}=|a_f|^2\ , \nonumber \\
A=W,Z\ &;& \quad \psi=b,\tau,c,\ldots \, 
\eea
where the $gg$ and $h$ final states are produced through a loop triangle diagram with only quarks as virtual particles. 

The calculation of the Higgs decay rate to two photons is more involved. By using the formulas given in \cite{Gunion:1989we}, we can write
\be\label{hgamgam}
\Gamma_{h\rightarrow \gamma\gamma}= \frac{\alpha_e^2 m_{h}^3}{256 \pi^3 v_w^2}\left| \sum_i  N_i e^2_i F_{i} \right|^2,
\ee
where the index $i$ is summed over the SM charged particles as well as $S^\pm$ and $W^{\prime\pm}$, $N_i$ is the number of colors, $e_i$ the electric charge in units of the electron charge, and the factors $F_{i}$ are defined by
\bea
F_{A}&=&\left[2+3 \tau_{A}+3 \tau_{A}\left( 2-\tau_{A} \right) f(\tau_{A})\right] a_V\ ,\quad A=W,W'\,;\nonumber\\
F_{\psi}&=&-2 \tau_{\psi}\left[1+\left( 1-\tau_{\psi} \right) f(\tau_{\psi})\right] a_f\ ,\quad \psi=t,b,\tau,\ldots\ ;\nonumber\\
F_{S}&=&\tau_{S}\left[ 1-\tau_{S}  f(\tau_{S})\right] a_S ,\ \, \tau_{i}=\frac{4 m_i^2}{m_{h}^2}\,,
\eea
with
\bea
f(\tau_{i})=\left\{
\begin{array}{ll}  \displaystyle
\arcsin^2\sqrt{1/\tau_{i}} & \tau_{i}\geq 1 \\
\displaystyle -\frac{1}{4}\left[ \log\frac{1+\sqrt{1-\tau_{i}}}
{1-\sqrt{1-\tau_{i}}}-i\pi \right]^2 \hspace{0.5cm} & \tau_{i}<1
\end{array} \right. .
\label{eq:ftau}
\eea
In the limit of heavy $W^{\prime\pm}$ and $S^\pm$, one finds
\be
F_{W'}=7\ ,\quad F_S=-\frac{1}{3}.
\label{fwps}\ee
Given the experimental lower bounds on $m_{W'}$ and $m_{S}$ \cite{ATLASWl,Heister:2002ev}, the error on $F_{W'}$ is irrelevant while $|F_S|$ gets enhanced by about $10\%$ for $m_S=150$ GeV: since the experimental error on $\hat{\mu}_{\gamma\gamma}$ is large and constructive interference of the $S^\pm$ and $W^\pm$ is favored by the experiment, we also assume the error involved by the above approximation for $F_S$ to be negligible. 

We notice also that in the limit of heavy masses for the charged scalar and vector bosons the light Higgs decay to such (virtual) states is highly suppressed by kinematics, and therefore no additional decay channels have to be taken into account besides those of the SM.

To evaluate the theoretical predictions for the measured observables, we need the SM production cross sections for the Higgs boson and the SM branching ratios for its decay. The production cross sections at the LHC and Tevatron for the final state $\Omega$ are given \cite{Dittmaier:2011ti} in Table~\ref{proddata}.
\begin{table}[htb]
\begin{tabular}{|c||c|c|c|c|c|c|}
\hline
$\Omega$ &$\, h\,$& $\, qqh\,$ & $\, t\bar{t}h\, $ &$\, Wh \,$ & $\, Zh \,$ &$\, h(X)\,$ \\
\hline
\, 7 TeV \,&  15.31 & 1.211 & 0.08634 & 0.5729 & 0.3158 & 17.50 \\
 \hline
\, 8 TeV \,& 19.52 & 1.578 & 0.1302 & 0.6966 & 0.3943 & 22.32\\
 \hline
\, 1 TeV \,& 0.9493 & 0.0653 & 0.0043 & 0.1295 & 0.0785 & 1.227 \\
\hline
\end{tabular}
\caption{Standard Model Higgs production cross sections in units of pb.}
\label{proddata}
\end{table}

The SM branching fractions are defined in terms of the decay rates, $\Gamma_{h\rightarrow ij}^{\textrm{SM}}$, as
\be
{\rm Br}^{\rm SM}_{ij}= \frac{\Gamma^{\rm SM}_{h\rightarrow ij}}{\Gamma^{\rm SM}_{\rm tot}}\ , \quad \Gamma^{\rm SM}_{\rm tot}=\sum_{ij=\bar{b}b,gg,WW,\ldots}\Gamma^{\rm SM}_{h\rightarrow ij}=4.03 {\rm MeV}.
\ee
These are given \cite{Dittmaier:2012vm} by
\bea
{\textrm{Br}}^{\rm SM}_{b\bar{b}} &=& 0.578, \,\,\,\,\quad {\textrm{Br}}^{\rm SM}_{\tau\bar{\tau}} = 0.0637,\,\,\qquad
{\textrm{Br}}^{\rm SM}_{c\bar{c}} = 0.0268, \quad {\textrm{Br}}^{\rm SM}_{gg} = 0.0856, \\
{\textrm{Br}}^{\rm SM}_{\gamma\gamma} &=& 0.0023, \quad {\textrm{Br}}^{\rm SM}_{\gamma Z} = 0.00155,\quad
{\textrm{Br}}^{\rm SM}_{WW} = 0.216, \quad {\textrm{Br}}^{\rm SM}_{ZZ} = 0.0267.
\eea

To determine the experimentally favored values of the free parameters $a_f,a_V,a_{V'},a_S$, we minimize the quantity
\be
\chi^2=\sum_i\left(\frac{{\cal O}_i^{\textrm{exp}}-{\cal O}_i^{\textrm{th}}}{\sigma_i^{\textrm{exp}}}\right)^2,
\label{chi2}\ee
where the measured values and errors of the observables are given in Tables~\ref{datatable},\ref{ggjjdata}, while the numerical predictions of the theory are easily determined from Eqs.~(\ref{hatgs}-\ref{fwps}), with the SM input values given in Table \ref{proddata}. In defining $\chi^2$ above, we assumed the correlation matrix to be simply the identity matrix.  
We note that it is not possible to constrain both $a_S$ and $a_{V'}$ since they both contribute only to the diphoton decay. The optimal values given below then refer to either of the two taken equal to zero:
\bea\label{opta}
a_V=0.97^{+0.10}_{-0.11}\ ,&\quad& a_f=1.02^{+0.25}_{-0.32}, \quad 
\left\{\begin{array}{c} a_{V'}=0.21^{+0.16}_{-0.18}\,\,\,{\textrm{and}}\, \ a_S=0 \\
a_{V'}=0\,\,\,{\textrm{and}}\,\, a_S=-4.4^{+3.8}_{-3.3}\end{array}\right.\ ,
\eea
with
\be\label{effLfit}
\chi^2_{\textrm{min}}/{\textrm{d.o.f.}}=0.85\ ,\quad P\left(\chi^2>\chi_{\textrm{min}}^2\right)=62\%\ ,\quad {\textrm{d.o.f}}.=14.
\ee
The probability to get a minimum value of $\chi^2$ larger than the optimal value above is naively expected to be around 50\%, and therefore the corresponding value obtained above shows that the simple parametrization of Eq.~(\ref{efflagr}) fits satisfactorily the data. 

As a comparison, the SM produces
\be\label{SMfit}
\chi^2_{\textrm{min}}/{\textrm{d.o.f}}.=0.92\ ,\quad P\left(\chi^2>\chi_{\textrm{min}}^2\right)=55\%\ ,\quad {\textrm{d.o.f.}}=17\ ,
\ee
for only the Higgs physics data, which indeed is a rather ideal result. The inclusion of the EW parameters $S$ and $T$ ($S=T=0$ for SM) \cite{Peskin:1991sw,He:2001tp,PDG} in the fit improves further the quality of the fit:
\be\label{SMfitST}
\chi^2_{\min}/{\textrm{d.o.f.}}=0.89\ ,\quad P\left(\chi^2>\chi_{\min}^2\right)=60\%\ ,\quad {\textrm{d.o.f.}}=19,
\ee
which shows that the SM is still perfectly viable in light of current collider data.

In the next section we use the Higgs physics constraints derived here to test the viability of a simple bosonic walking technicolor model \cite{Antola:2009wq} whose low energy effective Lagrangian belongs to the class specified by the generic Lagrangian of Eq.~(\ref{efflagr}). 

\section{Model and constraints}
\label{2HDMfit}

In Technicolor (TC) an additional, confining gauge interaction causes techniquarks, charged under TC and the EW interaction, to condense and break spontaneously the EW symmetry  \cite{Dimopoulos:1979es,Farhi:1980xs}. This mechanism allows the $W$ and $Z$ bosons, and the composite states of the strongly coupled TC sector to acquire mass, while the SM fermions remain massless.

The TC sector we consider has SU(2)$_L\times $SU(2)$_R$ chiral symmetry and is described in terms of the complex composite meson field $M^T=(\phi^+,\phi^0)$  by an effective Lagrangian
\be
{\cal L}_{\textrm{TC}}=D_\mu M^\dagger D^\mu M-m_M^2M^\dagger M-\frac{\lambda_M}{3!}\left(M^\dagger M\right)^2.
\ee
To provide mass for ordinary matter fermions, an additional interaction linking them to the TC condensate has to be provided.  In bosonic TC models, this link is provided by one (or more) elementary scalar(s) \cite{Simmons:1988fu,Samuel:1990dq,Kagan:1991gh,Carone:1992rh,Hemmige:2001vq}. While in this paper we consider nonsupersymmetric theories, bosonic technicolor effective Lagrangians also arise as low energy realizations of supersymmetric technicolor theories \cite{Dine:1981za, Antola:2010nt, Antola:2011bx}. In the context of bosonic TC, it is therefore the techniquark condensate that breaks EW symmetry, while the scalar plays the role of "spectator".  The Higgs Lagrangian is written in terms of the usual complex doublet $H$ as 
\be
{\cal L}_{\textrm{Higgs}}=D_\mu H^\dagger D^\mu H-m_H^2H^\dagger H-\frac{\lambda_H}{3!}\left(H^\dagger H\right)^2.
\label{higgsL}
\ee
The link between the technicolor and the SM matter fields obtained at the effective Lagrangian level is due to the Yukawa couplings of the Higgs field $H$. In addition to the usual couplings to the SM matter fields,
\be
{\cal L}_{\textrm{Yuk}}= (y_u)_{ij}  H \bar Q_i U_j + (y_d)_{ij} H^\dagger \bar Q_i D_j + (y_\ell)_{ij} H^\dagger \bar L_i E_j  + {\rm h.c.}\ ,
\ee
these include also the couplings to techniquarks, $y_{TC}\bar{\Psi}_L H \Psi_R$. When constructing the effective Lagrangian for the composite sector of the theory, this coupling generates further terms in the effective TC Lagrangian so that the technicolor sector is described by \cite{Antola:2009wq}\footnote{Compared to the potential presented in \cite{Antola:2009wq}, expressed in terms of matrix fields rather than EW doublets, we absorbed the $\omega$ factors in the $c_i$ coefficients and pulled out a factor $\lambda_H$ in front of $c_4$, as suggested by naive dimensional analysis.}
\bea
{\cal L}_{\textrm{bTC}} &=& D_\mu M^\dagger D^\mu M- m_M^2M^\dagger M-\frac{\lambda_M}{3!}\left(M^\dagger M\right)^2\nonumber \\
&+&\left[c_3 y_{TC} D_\mu M^\dagger D^\mu H+c_1 y_{TC} f^2 M^\dagger H+\frac{c_2y_{TC}}{3!} (M^\dagger M)(M^\dagger H)\right. \nonumber \\
&+&\left.\frac{c_4 y_{TC}}{3!}\lambda_H(H^\dagger H)(M^\dagger H)+{\textrm{h.c.}}\right]\ ,
\label{effL}
\eea
where $c_i$ are unknown parameters and $f$ is the vev of $M$.
The model that we consider is therefore specified by the effective Lagrangian
\be
{\cal L}={\cal L}_{\textrm{SM}}+{\cal L}_{\textrm{bTC}},
\label{fullLagr}
\ee
where ${\cal L}_{\textrm{SM}}$ is the usual SM Lagrangian containing the sectors 
${\cal L}_{\textrm{Higgs}}$ and ${\cal L}_{\textrm{Yuk}}$.
The coefficients $c_i$ in Eq.~(\ref{effL}) are estimated by naive dimensional analysis
\cite{Manohar:1983md,Cohen:1997rt} to be
\be\label{cNDA}
c_1\sim \omega\ ,\quad c_2\sim\omega\ ,\quad c_3\sim \omega^{-1}\ ,\quad c_4\sim \omega^{-1}\ ; \quad \omega\lesssim 4 \pi\ .
\ee 
Two of the parameters on the r.h.s. of Eqs.~(\ref{higgsL},\ref{effL}) are determined by the extremum conditions of the potential.
Furthermore, the electroweak scale constrains the vevs of $M$ and $H$ by
\be\label{vfvw}
v_w^2=v^2+f^2+2 c_3 y_{TC} f v=(246\ {\rm GeV})^2\ ,\quad \langle M \rangle=\frac{f}{\sqrt{2}}\ ,\quad \langle H \rangle=\frac{v}{\sqrt{2}}\,.
\ee
Finally, the requirement for the potential to be bounded from below imposes
\be\label{Vb}
\lambda_H,\lambda_M>0\ ; \quad \lambda_H+\lambda_M>2\left(c_2+c_4\lambda_H\right) y_{TC} .
\ee

The mass eigenstates are obtained by diagonalizing first the kinetic terms and then applying a rotation to diagonalize the mass terms in the scalar and pseudoscalar sectors. First, the Higgs fields $M$ and $H$ are expressed as
\be
\begin{pmatrix}
M \\
H
\end{pmatrix}=\frac{1}{\sqrt{2}}\begin{pmatrix} A & B \\ -A & B\end{pmatrix}\begin{pmatrix} M_2 \\ M_1\end{pmatrix},\quad
A=(1-c_3y_{TC})^{-1/2},\,\, B=(1+c_3y_{TC})^{-1/2}.
\label{cantr}\ee
After this transformation, the fields $M_{1,2}$ are written in terms of the charge eigenstates as
\be
M_{1,2} =\begin{pmatrix} \Sigma_{1,2}^\pm \\ \frac{1}{\sqrt{2}}(f_{1,2}+\sigma_{1,2}+i\xi_{1,2})\end{pmatrix}.
\ee
The rotation angles $\alpha$ and $\beta$ determine the physical states in the scalar and pseudoscalar sectors so that the Goldstone bosons, $G^\pm$ and $G^0$, provide the respective longitudinal components of the $W^\pm$ and $Z$ bosons, while $h^0,H^0,A^0$, and $H^\pm$ are the neutral scalars, pseudoscalar, and charged scalar mass eigenstates, respectively: 
\be\label{rotM}
\begin{pmatrix} h^0 \\ H^0 \end{pmatrix} = \begin{pmatrix} c_\alpha & -s_\alpha \\ s_\alpha & c_\alpha \end{pmatrix} \begin{pmatrix} \sigma_2 \\ \sigma_1 \end{pmatrix} ~, \
\begin{pmatrix} G^0 \\ A^0 \end{pmatrix} = \begin{pmatrix} s_\beta & c_\beta \\ c_\beta & -s_\beta \end{pmatrix} \begin{pmatrix} \xi_2 \\ \xi_1 \end{pmatrix} ~,\ 
\begin{pmatrix} G^\pm \\ H^\pm \end{pmatrix} = \begin{pmatrix} s_\beta & c_\beta \\ c_\beta & -s_\beta \end{pmatrix} \begin{pmatrix} \Sigma_2^\pm \\ \Sigma_1^\pm \end{pmatrix} ~.
\ee
The mixing angle $\beta$ is defined so that $\tan\beta=f_2/f_1$. The masses of the lightest composite states, including neutral scalars, are naturally expected to be of $O(\Lambda_{TC})\sim 1$ TeV. For bosonic TC the strong dynamics effect on the Higgs mass can be somewhat tamed by the mixing of the composite and elementary neutral scalar states, since the latter state can have a squared mass term much smaller than the former. This mechanism is analogous to the TeV-scale seesaw recently put forward in \cite{Foadi:2012ks}. Moreover, it has been shown \cite{Foadi:2012bb} that the top-quark loop contribution can greatly reduce the tree level TC prediction on the Higgs mass. A further suppression of the light Higgs mass is expected in NMWT because of walking dynamics \cite{Dietrich:2005jn}. From here on we assume that one or a combination of the mechanisms above is at work and use $m_{h^0}=125\pm 1$ GeV as an input to fix the value of one of the free parameters of the low energy effective theory.

To compare the model predictions with the LHC and Tevatron measurements, we need the coefficients of the SM Higgs linear couplings introduced in Eq.~\eqref{efflagr} to be expressed in terms of the bNMWT parameters:
\bea\label{afVS}
a_S&=&\left[\left(c_{2 \beta }-c_{2 \rho }\right) \left(\left(c_2-c_4\lambda_H\right) c^{-1}_{\rho } s^{-1}_{\rho } \left(c_{\alpha +3 \beta }+c_{\alpha -\beta } c_{2 \beta } c_{2 \rho }\right)\right.\right.\nonumber\\
&~&\left.\left.\qquad\qquad\quad\  +4
   \left(c_2+c_4\lambda_H\right) c_{\beta } s_{\beta } \left(c_{\alpha } c_{\beta } t_{\rho }^{-2}+s_{\alpha } s_{\beta } t_{\rho }^2\right)\right)\right.\nonumber\\
   &~&\left. \ \ -\left(c_{\alpha -\rho } s_{2 (\beta -\rho )}^2 s_{\beta +\rho } \lambda _H+c_{\alpha +\rho } s_{\beta -\rho } s_{2 (\beta +\rho )}^2 \lambda _M\right) c_{\rho }^{-2} s_{\rho }^{-2}/y_{TC}\right]\nonumber\\
&~&/\left[4 \left(c_4\lambda_H s_{\beta -\rho }^2+\left(12 c_1+c_2\right) s_{\beta +\rho }^2\right) \right]\ ,\nonumber\\
a_V&=&s_{\beta - \alpha}\ ,\quad a_f=\frac{c_{\alpha -\rho }}{s_{\beta -\rho }}\ ,\ 
\eea
where $s_\alpha,c_\alpha,t_\alpha$ are shorthands for $\sin\alpha,\cos\alpha,\tan\alpha$, respectively, with $\alpha,\beta$ defined by the rotation matrices in Eq.~\eqref{rotM} and $\rho$ by
\be
s_{\rho }=\sqrt{\frac{ 1 - c_3 y_{TC}}{2}}\ ,\quad c_{\rho }=\sqrt{\frac{ 1+c_3 y_{TC}}{2}}\ .
\ee
We are now ready to test the particle spectrum and its couplings against the latest experimental data. First, we scan the parameter space looking for data points that produce the right SM mass spectrum and satisfy the direct searches for charged particles at LEP \cite{Heister:2002ev} and a heavy neutral scalar at LHC \cite{Chatrchyan:2012tx} as well as the EW precision constraints from the $S$ and $T$ parameters \cite{Peskin:1991sw,He:2001tp,PDG}. More specifically, we impose the constraints
\bea\label{PHc}
m_{h^0}&=&125\pm 1\ {\rm GeV}\ ,\  m_{H^\pm}=m_{A^0}>100\ {\rm GeV}\ ,\  m_{H^0}>600\ {\rm GeV}\ ,\ |\frac{s_{\alpha -\rho }}{s_{\beta -\rho }}|<1\ ,\nonumber \\  
S&=&0.04\pm 0.09\ ,\ T=0.07\pm 0.08\ ,\ r(S,T)=88\%\ ,\ m_{A^0},m_{H^0}< 5 \Lambda_{TC}\ .
\eea
The quantity $r(S,T)$ is the correlation coefficient for the $S$ and $T$ parameters \cite{PDG}. The constraint on the trigonometric functions is to ensure that the heavy Higgs does not couple to SM fermions more strongly than a SM Higgs with the same mass does;  this allows us to use straightforwardly the LHC constraint on $m_{H^0}$. The upper bounds on $m_{A^0},m_{H^0}$ are enforced by the cutoff of $O(\Lambda_{TC}\approx4\pi v_w)$ of the effective Lagrangian. We also require the free parameters to produce the remaining SM mass spectrum and satisfy the bounds in Eq.~\eqref{Vb}. Then, we scan for such viable points in the domain
\bea\label{THc}
& &0<\lambda_H,\lambda_M<(4\pi)^2\ ,\ 2\pi<|c_1|,|c_2|,|c_3^{-1} |,|c_4^{-1}|<8\pi\ ,\ |c_3 y_{TC}|<1\nonumber\\
& &|y_t|<4\pi\ ,\ f=\pm \sqrt{v^2_w-v^2 \left(1-c_3^2 y_{TC}^2\right)}-v c_3 y_{TC}\ ,  |v|< v_w(1-c_3^2y_{TC}^2)^{-1/2}\ ,
\eea
with $m_H^2,m_M^2$ determined by the extremum conditions
\be\label{THc2}
\frac{\partial V}{\partial h^0}=0\ ,\quad\frac{\partial V}{\partial H^0}=0,
\ee
where $V$ is the scalar potential of the effective Lagrangian in Eqs.~(\ref{higgsL},\ref{effL},\ref{fullLagr}).
The results that we present in the following of this section can be applied directly to the Type-I 2HDM by using the formulas in Appendix \ref{2hdm}. The disclaimer is that we are testing only a portion of the parameter space available to such a model, and more precisely the range of values typical for underlying strong dynamics. 
 
\begin{figure}[htb]
\includegraphics[width=0.6\textwidth]{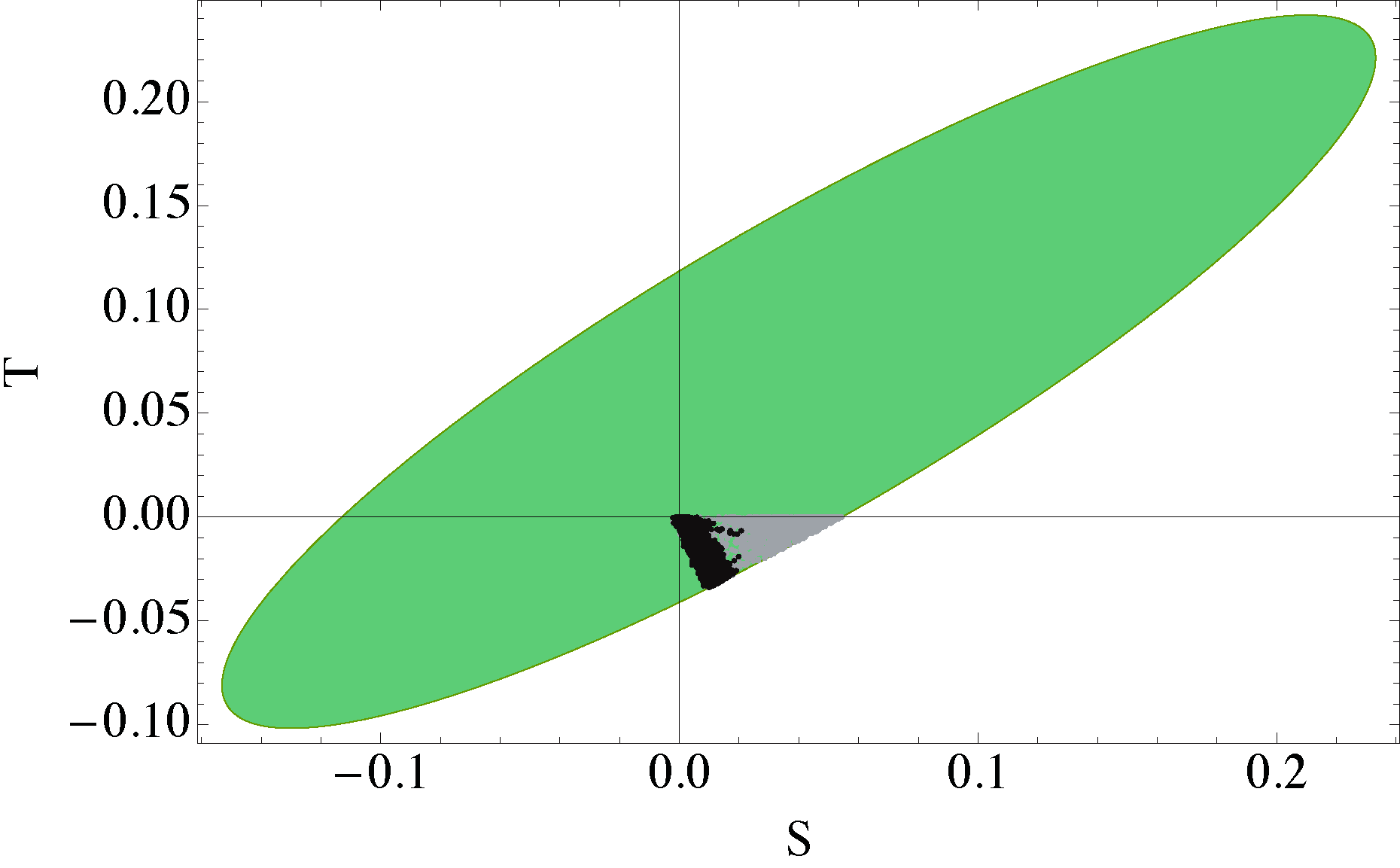}\hspace{0.5cm}
\caption{90\%CL viable region (in green) of the precision EW parameters $S$ and $T$: in black are the values relevant for bNMWT, while those in grey refer generically to Type-I 2HDM.}
\label{STp}
\end{figure}

The distribution of 5000 viable data points of the bNMWT allowed parameter space in the $(S,T)$ plane  is shown in Fig.~\ref{STp}. The 90\% Confidence Level (CL) allowed region is shaded in green while the viable data points featuring $m_H^2>0$ ($m_H^2<0$) are plotted in black (grey). We make this distinction because for positive $m_H^2$ the SM Higgs sector alone would not break EW symmetry, and therefore EW symmetry breaking is generated through bosonic TC interactions. The black dots are, thus, relevant for bNMWT while the grey ones refer more generically to the Type-I 2HDM.
It is clear that the viable region in $S$ and $T$ accessible by bNMWT is very limited.

\begin{figure}[htb]
\includegraphics[width=0.48\textwidth]{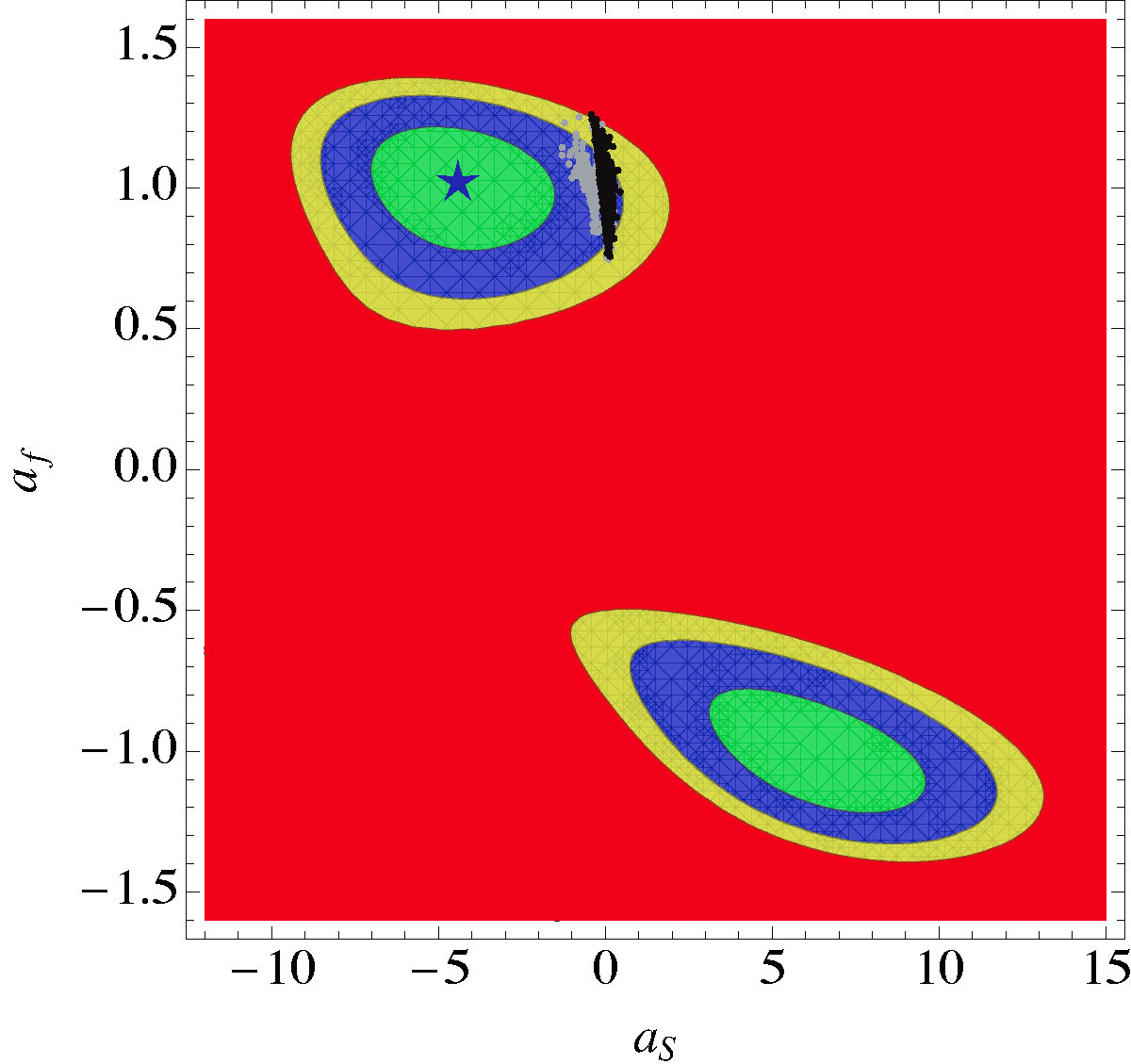}\hspace{0.45cm}
\includegraphics[width=0.48\textwidth]{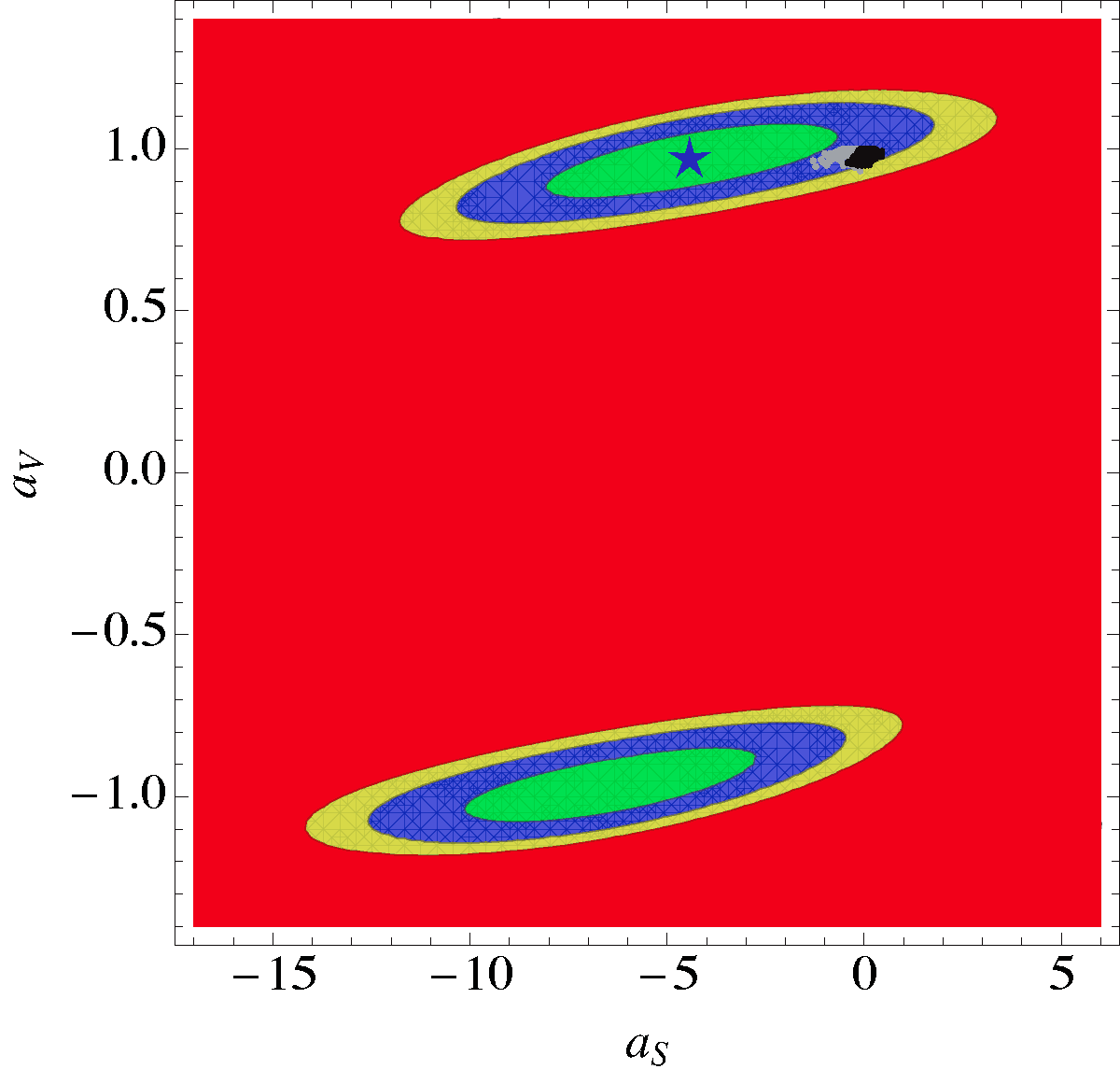}\hspace{0.45cm}
\caption{Viable data points in the $(a_S,a_f)$ (left pane) and $(a_S,a_V)$ (right pane) planes, together with the 68\% (green), 90\% (blue), and 95\% (yellow) CL region: in black are the values relevant for bNMWT while those in grey refer generically to Type-I 2HDM. The blue stars mark the optimal signal strengths on the respective planes intersecting the optimal point with $a_{V'}=0$.}
\label{aSafV}
\end{figure}

In Fig.~\ref{aSafV} we plot the viable data points in the $(a_S,a_f)$ and $(a_S,a_V)$ planes, respectively, together with the 68\% (green), 90\% (blue), and 95\% (yellow) CL regions obtained in the previous section: these plots represent a slice of the $a_f,a_V$, and $a_S$ parameter space passing through the optimal point (blue star) given in Eq.~\eqref{opta}. There is a perfectly specular viable region, which we do not show here, intersecting another $\chi^2$ global minimum point, obtained by flipping the signs of $a_V,a_f$, and $a_S$ in Eq.~\eqref{opta}.
 In Fig.~\ref{aSafV}, left panel, the upper viable region, containing the best fit point, obtains the observed slight enhancement of the Higgs diphoton decay rate exclusively by the charged scalar contribution which interferes constructively with the $W$ boson contribution. This results from a rather large linear coupling of $h^0$ to 
 $S^\pm$. In the lower viable region, the slight enhancement of the Higgs diphoton decay rate is entirely due to the SM fermions which couple to $h^0$ with the same sign as $W^\pm$ and, therefore, give constructive interference, while the scalar interferes destructively to balance an otherwise excessive enhancement of the decay rate to two photons. The same comments apply to the viable regions presented in Fig.~\ref{aSafV}, right panel.

The bNMWT data points are closely clustered around the SM values. This was somewhat expected, because of the small mixing of the two neutral Higgs fields for a heavy $H$ mass. The choice of a heavy masses for the new states is naively dictated by strong dynamics, which in the scaled up QCD case would predict masses of $O(\Lambda_{TC})\approx$ TeV. 

\begin{figure}[htb]
\includegraphics[width=0.55\textwidth]{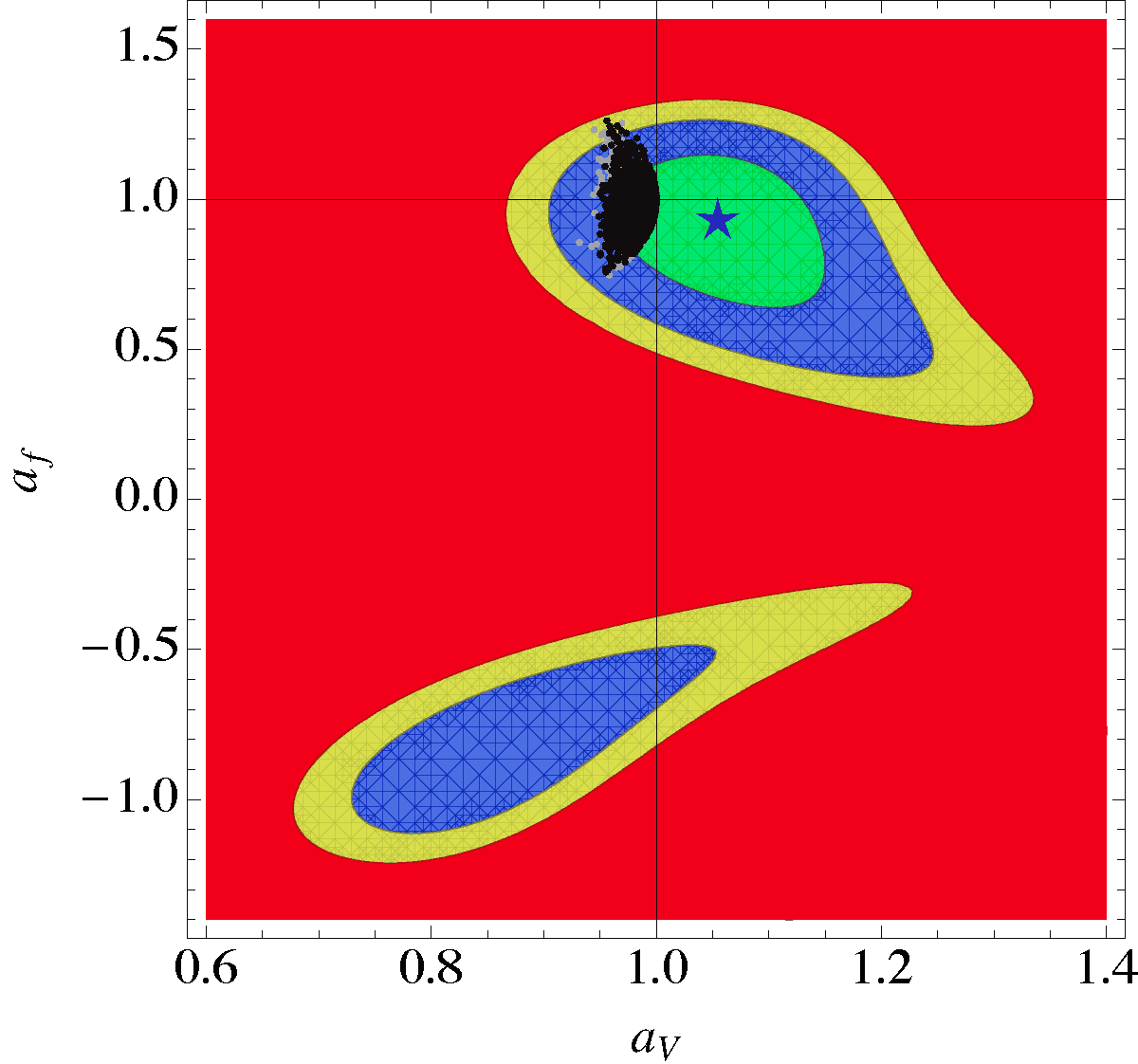}\hspace{0.5cm}
\caption{Viable data points in the $(a_V,a_f)$ plane, together with the 68\% (green), 90\% (blue), and 95\% (yellow) CL region: in black are the values relevant for bNMWT while those in grey refer generically to Type-I 2HDM. The blue star marks the optimal signal strengths on the $(a_V,a_f)$ plane for $a_S=a_{V'}=0$.}
\label{aVafSM}
\end{figure}

Finally, in Fig.~\ref{aVafSM} we show the bNMWT data points in the $(a_V,a_f)$ plane that passes through the SM point ($a_S=a_f=a_V=0$) to which they approximately belong. In bNMWT the Higgs-vector boson coupling is always reduced compared to its SM value, as shown in Fig.~\ref{aVafSM}, while the experiment favors an enhancement of the same coupling. While bNMWT looks generally disfavored compared to the SM, most of the scanned data points lie within the 90\% CL region.

We note that Fig.~\ref{STp} also includes bNMWT points with flipped sign of $a_f,a_V$, and $a_S$. These points belong to the plane passing through the specular optimal point, and therefore we do not include them in Figs.~\ref{aSafV} and~\ref{aVafSM}.

 Among the scanned 5000 viable bNMWT data points featuring $m_H^2>0$, the one minimizing $\chi^2$ is 
\be\label{minchib}
a_V=1.00\ ,\ a_f=0.98\ ,\ a_S=0.0\ ,\ \quad S=T=0\quad \Rightarrow\quad \chi^2=16.73\ .
\ee

To estimate the goodness of fit of the scanned data points, we have to determine the number of degrees of freedom (d.o.f.) of the bNMWT parameter space, limited by the constraints motivated by strong dynamics. Since the free variables in the bNMWT Lagrangian in Eqs.~(\ref{higgsL},\ref{effL}) can be grouped in the coupling coefficients of Eq.~(\ref{efflagr}), defined in terms of the TC Lagrangian variables by Eq.~(\ref{afVS}), one can expect the Higgs physics data presented in Section~\ref{datafit} to allow at most for three d.o.f., plus two since we also include the EW parameters $S$ and $T$ in the fit. In practice, the range of values found with the scan for $a_S,a_V,S,T$, is very small compared to the uncertainty affecting each of those parameters, and becomes negligible in the neighborhood of the best fit value of $a_f$ due to a high correlation among all the five parameters. We assume, therefore, to have only one free parameter, $a_f$, which produces the following statistical results:
\be\label{STfitbNMWT}
\chi^2_{\min}/\textrm{d.o.f.}=0.93\ ,\quad P\left(\chi^2>\chi_{\min}^2\right)=54\%\ ,\quad \textrm{d.o.f.}=18\ .
\ee
These numbers should be compared to the corresponding results in the SM, Eq.~\eqref{SMfitST}, which indeed produces a better fit. Even without including $S$ and $T$ in the fit, we obtain the same optimal data point as the one in Eq.~(\ref{minchib}), with goodness of fit given by:
\bea
\chi^2_{\min}/\textrm{d.o.f.}&=&0.97\ ,\quad P\left(\chi^2>\chi_{\min}^2\right)=49\%\ ,\quad \textrm{d.o.f.}=16\ .
\eea
This result again is statistically worse than the relevant SM result, Eq.~\eqref{SMfit}, and less appealing than the general fit result,  Eq.~\eqref{effLfit}.

For further comparison, we also give the corresponding results for the generic 2HDM data points ($m_H^2<0$):
\bea
a_V&=&0.99\ ,\quad a_f=0.93\ ,\quad a_S=-1.2\ ;\nonumber\\ 
\chi^2_{\min}/\textrm{d.o.f.}&=&0.87\ ,\quad P\left(\chi^2>\chi_{\min}^2\right)=60\%\ ,\quad \textrm{d.o.f.}=16.
\eea
After adding the $S$ and $T$ parameters as well to the fit, we obtain
\bea\label{minchi2HD}
a_V&=&1.00\ ,\quad a_f=0.99\ ,\quad a_S=-0.4\ ,\ \quad S=0.01\ , \quad T=0\ ;\nonumber\\ 
\chi^2_{\min}/\textrm{d.o.f.}&=&0.91\ ,\quad P\left(\chi^2>\chi_{\min}^2\right)=57\%\ ,\quad \textrm{d.o.f.}=18.
\eea
Indeed, the data points featuring $m_H^2<0$ produce a better fit than that of the bNMWT with EW symmetry breaking determined by strong dynamics, Eq.~(\ref{STfitbNMWT}). 

However, the results obtained so far for bNMWT do not take into account the contributions to Higgs physics coming from heavy charged vector bosons, which are a staple of strong dynamics. In the next section we study this subject by introducing some simple interaction terms in the Lagrangian and by working out the corresponding Higgs physics phenomenology.

\section{Extra Charged Vector Bosons and Experimental Data Fit}
\label{Wpfit}
Extra vector bosons arise naturally in TC as composite resonances with a mass of the order of the strong interaction scale $\Lambda_{TC}$. 
Of particular interest to us here is the possibility that an extra charged vector boson, $W'$, be responsible for the observed slight enhancement to the  diphoton decay rate of the Higgs even if $m_{W'}\gg m_{h^0}$. 
Given the LHC constraints on the mass $m_{W'}$, equal to 2.55 TeV \cite{ATLASWl}, it is safe to take the heavy $W'$ limit, Eq.~\eqref{fwps}, for the Higgs decay rate to $\gamma\gamma$ in Eq.~\eqref{hgamgam}. 

To introduce, in the effective Lagrangian, direct Higgs couplings to an extra massive vector boson which conserves gauge symmetry at the fundamental scale, we use the hidden local symmetry principle \cite{Bando:1984ej,Bando:1987br}, which has been already applied to NMWT in \cite{Foadi:2007ue,Belyaev:2008yj}: here we just outline the main steps required to introduce composite vector bosons while conserving gauge invariance in the fundamental theory.

We begin by defining the following covariant derivatives
\be
D^\mu N_L=\partial^\mu N_L+i g_L \tilde{W}^\mu N_L +i g_{TC} A_L^\mu N_L \ ,\quad D^\mu N_R=\partial^\mu N_R+i g_Y \tilde{B}^\mu N_R+i g_{TC} A_R^\mu N_R\ ,
\ee
where $A_\mu$ is the vector boson associated with the $G\equiv SU(2)\times SU(2)$ global symmetry of ${\cal L}_{TC}$, which we have gauged above, and $N_L$ $(N_R)$ is a scalar field in the fundamental of $SU(2)_L$ $(U(1)_Y)$ and antifundamental of $G$. From the equations above, we can define a new vector field and its transformation under the full gauged symmetry by
\be\label{compV}
{\rm Tr}\left[N_L N_L^\dagger\right] P_L^\mu=\frac{D^\mu N_L N_L^\dagger-N_L D^\mu N_L^\dagger}{i g_{TC}}\ ,\quad P_L^\mu\rightarrow u_L P_L^\mu u_L^\dagger\ ,
\ee
where $u_L$ is a unitary transformation operator of $SU(2)$. The definition and transformation law of $P_R$ are obtained simply by replacing $L$ with $R$ in Eq.~\eqref{compV}. Among the possible dimension four, gauge invariant $P^\mu$ coupling terms to $M$, we retain only the following
\be\label{VLag}
{\cal L}_{M-P}=-g^2_{TC} r_2 {\rm Tr} \left[P_{L\mu} M' P^\mu_R M^{\prime\dagger}\right]+\frac{g_{TC}^2 r_1}{4}{\rm Tr}\left[P_{L\mu}^2+P_{R\mu}^2\right] {\rm Tr} \left[M' M^{\prime\dagger}\right]\ ,
\ee
where $M'$ is the matrix representation of the EW doublet $M$. Assigning non-zero vevs for $N_L$ and $N_R$, their kinetic terms generate a squared mass term for two vector boson combinations:
\be\label{Vvev}
m_A^2 {\rm Tr}\left[C_{L\mu}^2+C_{R\mu}^2\right]\ ,\quad C_L^\mu\equiv \langle P_L^\mu\rangle=A_L^\mu-\frac{g_L}{g_{TC}}\tilde{W}^\mu\ ,\quad C_R^\mu\equiv \langle P_R^\mu\rangle=A_R^\mu-\frac{g_Y}{g_{TC}} \tilde{B}^\mu\ .
\ee
The resulting massless eigenstates give the ordinary $W^\mu$ and $B^\mu$ vector bosons, which instead acquire mass through EW symmetry breaking. In addition, there are two vector boson triplets, one vectorial ($V^\mu$) and the other axial ($A^\mu$). Since their interaction terms, given by Eq.~\eqref{VLag} evaluated at the vev defined in Eq.~(\ref{Vvev}), respect custodial symmetry and give the same contribution to the axial-axial and vector-vector EW vector boson polarization functions \cite{Peskin:1990zt}, the total contribution of the vector bosons to the EW precision parameters is identical to the SM one, and  $S$ and $T$ are, therefore, zero \cite{Foadi:2007ue,Belyaev:2008yj}. Moreover the vector boson contributions to FCNC have been tested against the experimental results and shown to be phenomenologically viable \cite{Fukano:2011is}.  To simplify our analysis we fix
\be
r_2=-r_1\ ,
\ee
so that only $\tilde{W}^\mu$ and the vector resonance, $V^\mu$, couple to the neutral Higgs fields. 

The charged vector boson mass matrix in the $(\tilde{W},V,A)$ basis can be written in a compact form as
\be
\left(
\begin{array}{ccc}
 m_{\tilde{W}}^2 & -\frac{\epsilon  m_V^2}{\sqrt{2}} & -\frac{\epsilon  m_A^2}{\sqrt{2}} \\
 -\frac{\epsilon  m_V^2}{\sqrt{2}} & m_V^2 & 0 \\
 -\frac{\epsilon  m_A^2}{\sqrt{2}} & 0 & m_A^2
\end{array}
\right)\ ,
\label{generalvectormasses}
\ee
with
\be
m_{\tilde{W}}=\left[x^2+\left(1+s^2\right) \epsilon ^2\right] m_A^2\ ,  \quad  m_V^2=\left(1+2 s^2\right) m_A^2\  ,
\label{svar2}\ee
and
\be
s\equiv\frac{g_{\text{TC}} f}{2 m_A}  \sqrt{r_1}\ ,\quad x\equiv \frac{g_L v_w}{2 m_A}\ ,\quad  \epsilon\ \equiv\frac{g_L}{g_{TC}}\ .
\label{svar}\ee

We now study the implications of this setup in light of the LHC and Tevatron data fit we have at our disposal.

\subsection{Mixing of Vector Fields}
Let us begin with a rather general and simple case. We require that only $\tilde{W}$, the elementary gauge field, couples to $h^0$, and therefore the $W'$ coupling to the light Higgs is generated only through terms mixing $\tilde{W}$ with the composite vector fields $V$ and $A$. The squared mass matrix in the gauge basis  $(\tilde{W},V,A)$ is obtained simply by setting $s=0$ in Eqs.~(\ref{generalvectormasses},\ref{svar2}):
\be\label{Wmatr}
\left(
\begin{array}{ccc}
 \frac{ g_L^2 v_w^2}{4}+\epsilon ^2 m_A^2 & -\frac{\epsilon  m_A^2}{\sqrt{2}} & -\frac{\epsilon  m_A^2}{\sqrt{2}} \\
 -\frac{\epsilon  m_A^2}{\sqrt{2}} & m_A^2 & 0 \\
 -\frac{\epsilon  m_A^2}{\sqrt{2}} & 0 & m_A^2
\end{array}
\right)\ .\ee
We define the rotation to the mass eigenbasis in terms of $x$ and $\epsilon$, Eqs.~(\ref{svar}), by
 \be\label{Wmbasis}
\left(
\begin{array}{c}
 \tilde{W} \\
 V \\
 A
\end{array}
\right)=\left(
\begin{array}{ccc}
 c_{\varphi } & -s_{\varphi } & 0 \\
 \frac{s_{\varphi }}{\sqrt{2}} & \frac{c_{\varphi }}{\sqrt{2}} & -\frac{1}{\sqrt{2}} \\
 \frac{s_{\varphi }}{\sqrt{2}} & \frac{c_{\varphi }}{\sqrt{2}} & \frac{1}{\sqrt{2}}
\end{array}
\right) \left(
\begin{array}{c}
 W \\
 W' \\
 W''
\end{array}
\right)\ ,\quad c_{\varphi }=\frac{1}{\sqrt{2}}\sqrt{1+\frac{1-x^2-\epsilon ^2}{\sqrt{\left(1+x^2+\epsilon ^2\right)^2-4 x^2}}}\ ,
 \ee
with corresponding eigenvalues
\be\label{Wpmass}
m^2_{W,W'}=\frac{1}{2} \left[1+x^2+\epsilon ^2\mp \sqrt{\left(1+x^2+\epsilon ^2\right)^2-4 x^2}\right] m_A^2\ ,\quad m^2_{W''}=m_A^2\ .
\ee

The mixing matrix in Eq.~(\ref{Wmbasis}) shows that only $W$ and $W'$ contribute to the gauge field $\tilde{W}$.

We checked that the Fermi coupling, $G_F$, determined by evaluating the amplitude for the muon decay ($\mu^-\rightarrow \nu_\mu\bar{\nu}_e e^-$), respects the usual relation
\be
\sqrt{2} G_F=v_w^{-2}=(246\,{\textrm{GeV}})^{-2}\ .
\ee

The vector coupling coefficient $a_V$ is suppressed, compared to the result in Eq.~(\ref{afVS}), because of  mixing:
\be
a_V=c^2_{\varphi '} s_{\beta - \alpha}\ ,\quad a_{V'}= s^2_{\varphi '} s_{\beta - \alpha}\ ,
\ee
with
\be\label{cphis}
c_{\varphi '}^2=\frac{g_L^2 v_w^2}{4 m_W^2} c_{\varphi }^2=\frac{2 x^2 \epsilon ^2}{\left(1+x^2+\epsilon ^2\right)^2-4 x^2-\left(1-x^2+\epsilon ^2\right)
   \sqrt{\left(1+x^2+\epsilon ^2\right)^2-4 x^2}}\ .
\ee
The $W''$ coupling to $h^0$ is zero instead because we set $r_2=-r_1$ in Eq.~\eqref{VLag}. Finally, the fermion and scalar coupling coefficients in Eq.~\eqref{afVS} remain unchanged.

The lower limit on the mass of a sequential\footnote{A sequential $W'$ has the same couplings as the SM $W$.} $W'$ from direct searches at ATLAS \cite{ATLASWl}, equal to 2.55 TeV at 95\%CL, can be readily applied to the case above by properly rescaling the lower limit:
\be
\Gamma_{W'\rightarrow l \nu}=\frac{g_{W'}^2 m_{W'}}{48 \pi} \quad \Rightarrow \quad  m_{W'}>(2.55 {\rm TeV})\left(\frac{g_{W'}}{g_W}\right)^2\ ,\ \frac{g_{W'}}{g_W}=-s_{\varphi '} \frac{m_{W'}}{m_W}\ ,
\label{ATWlb}\ee
with $s_{\varphi '},m_{W'},m_W$ defined in terms of $x,\epsilon$, and $m_A$ by Eqs.~(\ref{cphis},\ref{Wpmass}). By plotting the experimentally viable region defined by Eqs.~\eqref{ATWlb} on the $x$ and $\epsilon$ plane, we find the maximum allowed value of $\epsilon=0.36$ (at the 95\%CL), reached at $x=0$ (equivalent to the limit of large $m_A$).

In strong dynamics the value of $m_A$ is expected to be of $O(\rm TeV)$, which determines $x$ to be of  $O(10^{-2})$. In general this needs not to be the case, as larger values of $x$ for small $\epsilon$ are allowed by the experiment. On the other hand we are interested primarily in testing bNMWT, and therefore in the following we will limit our analysis by assuming $x\ll 1$. This choice, moreover, guarantees that the $W$ couplings do not change dramatically. Also, $\epsilon$ is expected to be small because of Eqs.~\eqref{svar} and the fact that $g_{TC}\gg g_L$. An expansion in both $\epsilon$ and $x$ therefore produces
\be
a_V=s_{\beta - \alpha}\left(1-x^2\epsilon^2\right)+O\left(x^{n}\epsilon^{5-n}\right)\ ,\quad a_{V'}=s_{\beta - \alpha}x^2\epsilon^2+O\left(x^{n}\epsilon^{5-n}\right)\ ,\quad n=0,\ldots,5\ .
\ee
In this limit the effect of mixing on the $W'$ and $W$ couplings is negligible. Moreover, because the sum of $a_V$ and $a_{V'}$ is independent of $\epsilon$, so is the Higgs decay rate to two photons.
Therefore, the optimal values for bNMWT with $m_A=1$ TeV, $\epsilon<0.36$, and $m_H^2>0$ ($m_H^2<0$) correspond to the ones with no mixing ($\epsilon=0$), presented in Eqs.~(\ref{minchib}) (Eqs.~(\ref{minchi2HD})).

In the next subsection we study the phenomenologically more appealing scenario in which the composite vector fields feature a direct coupling to neutral Higgs fields.

\subsection{Direct Higgs Coupling to $W'$ and $W''$}
Next we want to study the effects of a direct Higgs coupling to the composite vector field $V$, and consider $s\neq 0$. In this case, the charged vector boson mass matrix in the $(\tilde{W},V,A)$ basis is given by Eq.~(\ref{generalvectormasses}). 
The mass eigenvalues are lengthy cubic roots. These can be expanded in $x$ and $\epsilon$, which in TC are both expected to be small:
\bea
m_W^2 &\cong& m_A^2 x^2\left[1-\epsilon ^2\right]\ ,\quad m^2_{W''}\cong m_A^2 \left[1+\frac{1}{2} \left(1+x^2\right) \epsilon ^2-\frac{1}{8} \left(2+\frac{1}{s^2}\right) \epsilon ^4\right]\ ,\nonumber\\
m_{W'}^2 &\cong& m_A^2 \left[1+2 s^2+\frac{1}{2} \left(1+2 s^2+x^2\right) \epsilon ^2+\frac{1}{8} \left(2+\frac{1}{s^2}\right) \epsilon ^4\right]\ ,
\eea
where contributions of $O(x^n\epsilon^{5-n})$ are neglected, with $n=0,\ldots,5$.
The $\tilde{W}$ and $V$ coupling terms to the light Higgs can be derived from the mass matrix by taking its derivative with respect to $v_w$ and introducing a factor $\zeta$ to take into account the rotation of $M$ to the mass eigenbasis:
\be\label{hWW}
{\cal L}\subset \frac{2 m_A^2}{v_w} s_{\beta - \alpha} \left[\left(x^2+\zeta s^2 \epsilon ^2\right) \tilde{W}\tilde{W}+2 \zeta s^2 V V-2 \sqrt{2} \zeta s^2 \epsilon  \tilde{W} V\right] h^0\ ,\quad \zeta=s_{\beta-\alpha}^{-1}\frac{c_{\alpha+\rho}}{s_{\beta+\rho}}\ .
\ee
The vector boson coupling coefficients get an enhancement factor because of the $s$ coupling:
\be\label{directa}
a_V=\eta_W s_{\beta - \alpha}\ ,\quad a_{V'}= \left(\eta_{W'}+\eta_{W''}\right) s_{\beta - \alpha}\ ,
\ee
where
\bea\label{etaeq}
\eta_W&\cong& 1-\frac{\left[1+s^2 \left(3-\zeta\right)+2 s^4\right] x^2 \epsilon ^2}{\left(1+2 s^2\right)^2}\ ,\quad \eta_{W'}\cong\frac{2 \zeta s^2}{1+2 s^2}+\frac{\left[1+2 s^2\left(1-\zeta\right)\right]x^2 \epsilon ^2}{2 \left(1+2 s^2\right)^2}-\frac{\zeta\epsilon ^4}{8 s^2}\ ,\nonumber\\ \eta_{W''}&\cong&\frac{x^2 \epsilon ^2}{2}+\frac{\zeta\epsilon ^4}{8 s^2}\ ,
\eea
at  $O(x^n\epsilon^{5-n})$, with $n=0,\ldots,5$. We collected together the $W'$ and $W''$ contributions to the Higgs decay to diphoton by summing up their respective coupling coefficients in Eq.~\eqref{directa}. It is interesting to notice that, at all orders in $x$:
\be
\eta_W+\eta_{W'}+\eta_{W''}=1+\frac{2\zeta s^2}{1+2 s^2}+O(\epsilon^5)\ .
\ee
The fermion and scalar coefficients are still determined by Eqs.~\eqref{afVS}.
We obtain the optimal value of $s$ by performing the global fit in the limit of negligible vector mixing ($\epsilon=0$) and decoupled neutral heavy Higgs:
\be
a_f=a_V=1\ ,\quad a_S=0\ ,\quad a_{V'}=\frac{2 s^2}{1+2 s^2}\ ,\quad \Rightarrow \quad s =0.32^{+0.17}_{-0.32}\ .
\ee

\begin{figure}[htb]
\includegraphics[width=0.48\textwidth]{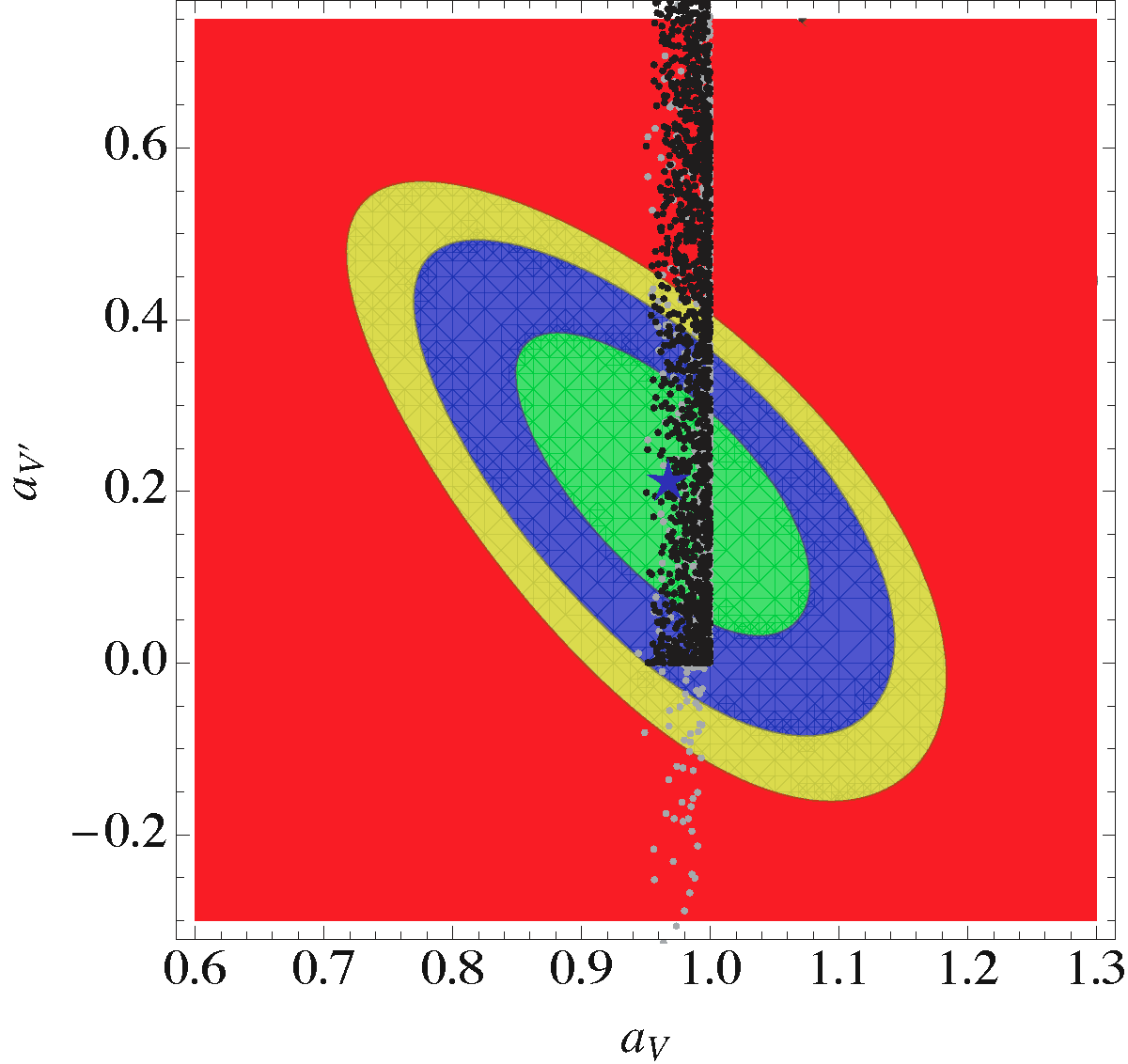}\hspace{0.45cm}
\includegraphics[width=0.48\textwidth]{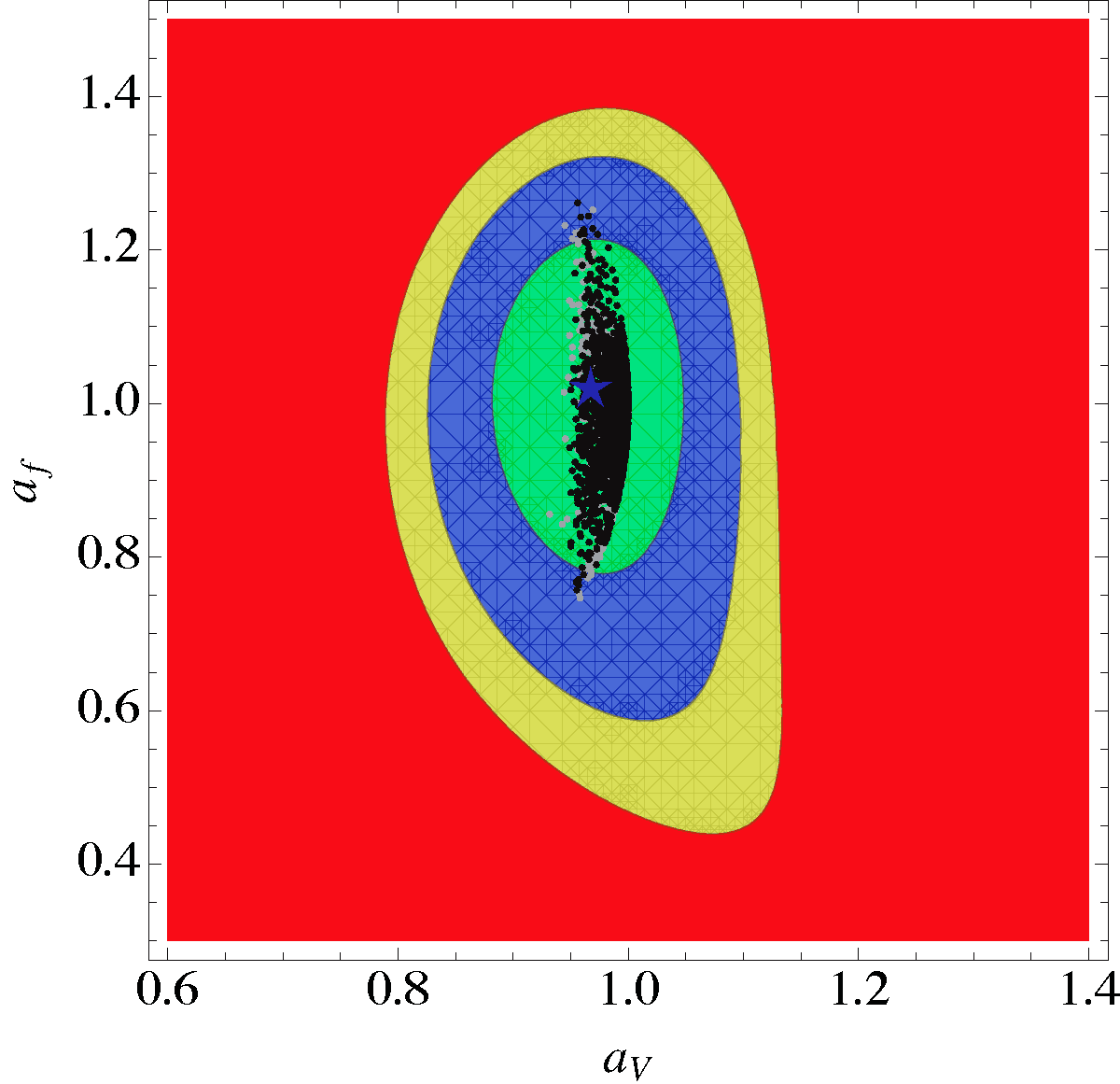}\hspace{0.45cm}
\caption{Viable data points in the $(a_V,a_{V'})$ and $(a_V,a_f)$ planes, together with the 68\% (green), 90\% (blue), and 95\% (yellow) CL region: in black are the values relevant for bNMWT while those in grey refer generically to Type-I 2HDM with the addition of two charged vector bosons. The blue stars mark the optimal signal strengths on the respective planes intersecting the optimal point with $a_{S}=0$.}
\label{aVVpf}
\end{figure}

\begin{figure}[htb]
\includegraphics[width=0.55\textwidth]{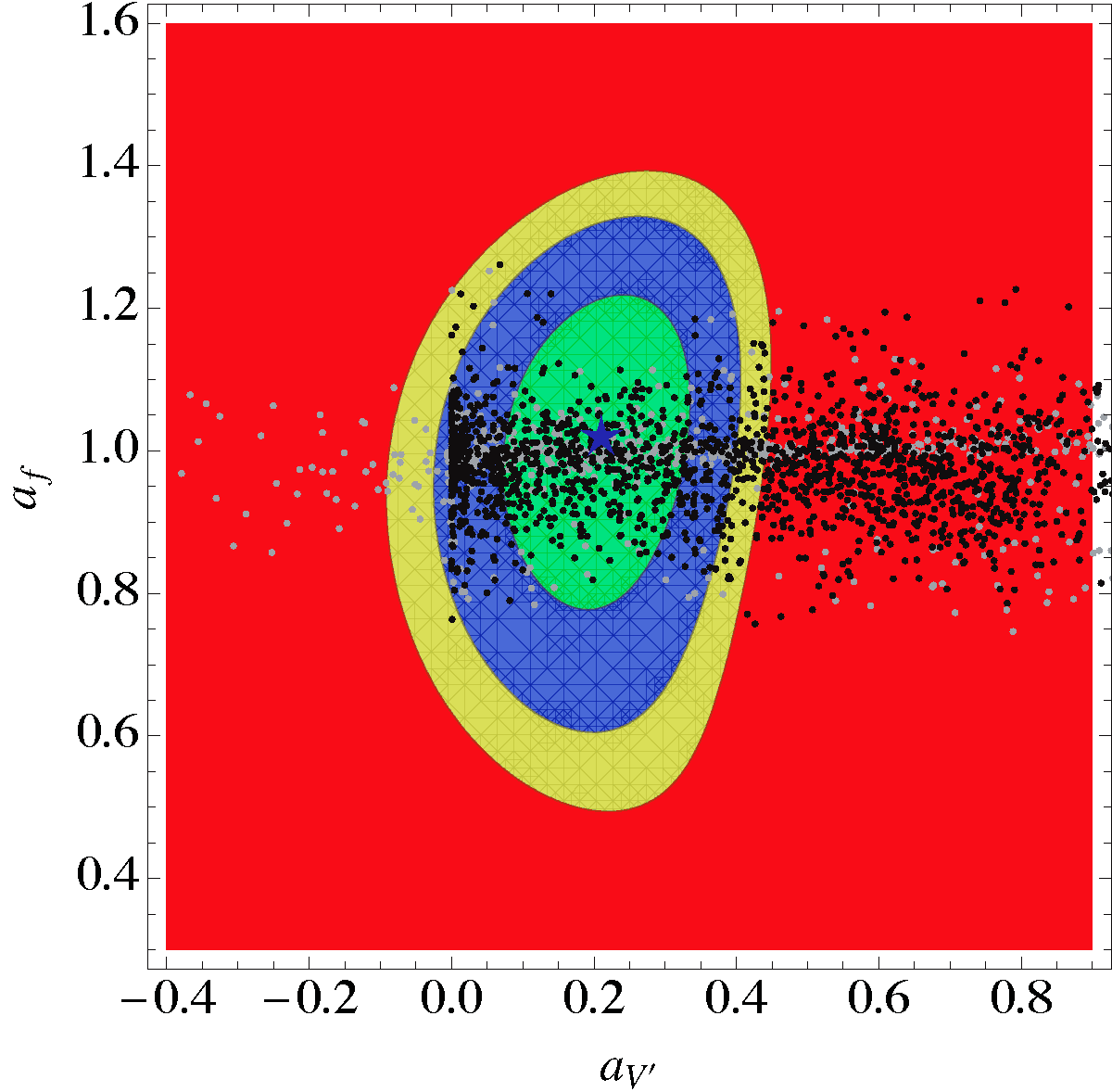}\hspace{0.5cm}
\caption{Viable data points in the $(a_{V'},a_f)$ plane, together with the 68\% (green), 90\% (blue), and 95\% (yellow) CL region: in black are the values relevant for bNMWT while those in grey refer generically to Type-I 2HDM with the addition of two charged vector bosons. The blue star marks the optimal signal strengths on the $(a_{V'},a_f)$ plane for optimal $a_V$ and with $a_{S}=0$.}
\label{aVpf}
\end{figure}

We use the same set of 5000 viable points scanned over the bNMWT parameter space with no $W'$ and $W''$, and re-calculate the coupling coefficients  $a_V$ and $a_{V'}$ at each data point for random values of $s$ and $\epsilon$, with $0\leq s\leq 1$ and $0\leq \epsilon \leq 0.1$, and $m_A=1$ TeV. We plot the resulting bNMWT data points together with the experimentally favored regions in the $(a_V,a_{V'})$ and $(a_V,a_f)$ planes in Fig.~\ref{aVVpf}, and in the $(a_{V'},a_f)$ plane in Fig~\ref{aVpf}, respectively, all passing through the optimal data point defined in Eq.~\eqref{opta}. The plots are again limited to the positive $a_V$ half-plane. For $m_H^2<0$ the mixing factor $\zeta$, Eq.~\eqref{hWW}, can be negative, which makes $a_{V'}$ flip sign, compared to $a_V$, because of Eqs.~\eqref{etaeq}.
Already by visual inspection, it is clear that the $s$ coupling allows bNMWT to cover a large portion of the 68\% CL favored region. We verified that all the three parameters $a_f,a_{V'}$, and (to a lesser degree) $a_V$ are free (meaning that they are little correlated and with a range of values comparable to the error affecting the optimal values in Eq.~\eqref{opta}), which was expected since we introduced two new parameters, $\epsilon$ and $s$. In this case the bNMWT data point minimizing $\chi^2$ for $m_H^2>0$ is
\bea\label{minchibTCs}
a_V&=&1.00\ ,\quad a_f=1.00\ ,\quad a_{V'}=0.19\ ,\quad a_{S}=0.0\ ,\quad S=T=0.00\ ;\nonumber\\ 
\chi^2_{\min}/\textrm{d.o.f.}&=&0.83\ ,\quad P\left(\chi^2>\chi_{\min}^2\right)=65\%\ ,\quad \textrm{d.o.f.}=16,
\eea
while for $m_H^2<0$ we find
\bea\label{minchi2HDs}
a_V&=&1.00\ ,\quad a_f=0.98\ ,\quad a_{V'}=0.18\ ,\quad a_{S}=0.0\ ,\ \quad S=T=0.00\ ;\nonumber\\ 
\chi^2_{\min}/\textrm{d.o.f.}&=&0.83\ ,\quad P\left(\chi^2>\chi_{\min}^2\right)=65\%\ ,\quad \textrm{d.o.f.}=16.
\eea
The cases above are equally favored by the experiment and both feature a probability greater than the one for the general fit, Eq.~(\ref{opta}), which does not include the $S$ and $T$ EW parameters (and therefore has two less d.o.f.) but gives a similar value of $\chi^2_{min}$. As a last remark, we note that $a_{V'}$ is rather unconstrained by the chosen set of observables, and so a broader set of observables related to $W'$ physics would be necessary to further test the viability of bNMWT.

\section{Conclusions and outlook}
\label{checkout}

In this paper we have considered quantitatively how much the coefficients of Higgs couplings to electroweak gauge bosons $(a_V)$ and fermions $(a_f)$ as well as to possible extra scalars $(a_S)$ can deviate from their corresponding values in the Standard Model $(a_V=a_f=1$, $a_S=0)$ in light of the current LHC and Tevatron data. We then considered a bosonic technicolor model, bNMWT \cite{Antola:2009wq}, and studied its viability by performing a scan on the parameter space which implements the direct constraints on the mass spectrum as well as the constraints from precision EW data (in terms of the $S$ and $T$ parameters). The scalar sector of the bosonic technicolor model we considered can be more generally viewed as a type-I 2HDM. 

The essential consequence of underlying strong dynamics is the existence of new vector resonances in the particle spectrum. We implemented these new states in an effective Lagrangian to study their effects. We considered in detail the implications of the effective Lagrangian on the couplings of the Higgs boson to the physical $W$ and $Z$ bosons as well as to fermions. Then, we studied first the simple case of minimal coupling to the SM fields, which amounts to considering only the mixing of the two new triplets of vector bosons and the SU(2)$_L$ gauge fields without direct interaction between the composite vector bosons and SM fields. In this simple scenario we determined the 95\%CL upper bound on the amount of mixing allowed by experimental data.  We showed that this generic scenario in bNMWT cannot be resolved by the current data. 

Finally we illustrated the possible effects of the direct coupling between composite vector bosons and neutral scalar fields within our effective Lagrangian scheme. We showed that the direct coupling allows for an optimal fit by the bNMWT predictions of the current experimental data. A more refined analysis of the model including additional observables can further test the possibility of a strongly coupled sector underlying the electroweak sector of the SM.

\acknowledgments
The support from Finnish Cultural Foundation, Central Finland Regional fund is gratefully acknowledged.

\newpage
\appendix

\section{Two Higgs Doublet Model Potential}\label{2HDMV}
\label{2hdm}
Let us write our bosonic technicolor Lagrangian explicitly as a two-Higgs doublet model (2HDM). The starting point is given by Eqs.(\ref{effL},\ref{fullLagr}). The kinetic mixing term in Eq.(\ref{effL}) is rotated away and the model canonically normalized by Eq.\eqref{cantr}. Applying one more rotation we can express the SM Yukawa couplings in the following form
\be\label{Yuk2HD}
{\cal L}_{\textrm{Yuk}} = (y_u)_{ij}  K H_2 \bar Q_i U_j + (y_d)_{ij} K H_2^\dagger \bar Q_i D_j + (y_\ell)_{ij} K H_2^\dagger \bar L_i E_j  + {\rm h.c.}\ ,\quad K=\left(1-c_3^2 y_{TC}^2\right)^{-\frac{1}{2}}\ ,
\ee
where the full transformation is given by
\be
H= K H_2\ ,\quad M=H_1-c_3 y_{TC} K H_2\ .
\label{2HDrot} \ee
As Eq.~(\ref{Yuk2HD}) shows, only one of the two Higgses (by convention $H_2$) couples to SM fermions, which ensures that there are no tree-level contributions to Flavor Changing Neutral Currents (FCNC): such a model in literature has been referred to as Type-I 2HDM \cite{Branco:2011iw}. 

On the other hand, the most general renormalizable Higgs potential of a 2HDM can be written as
\begin{eqnarray}\label{2HDMV}
V &=& m_{1}^2H_1^\dagger H_1 + m_{2}^2 H_2^\dagger H_2 - m^2_{12} \left(H_1^\dagger H_2+H_2^\dagger H_1\right) + \frac{\lambda_1}{2} (H_1^\dagger H_1)^2 + \frac{\lambda_2}{2} (H_2^\dagger H_2)^2 \nonumber \\
   &+& \lambda_3 (H_1^\dagger H_1)(H_2^\dagger H_2) + \lambda_4 (H_2^\dagger H_1)(H_1^\dagger H_2) +\left[ \frac{\lambda_5}{2} (H_1^\dagger H_2)^2 \right.  \nonumber \\
   &-& \left. \lambda_6 (H_2^\dagger H_1)(H_1^\dagger H_1) - \lambda_7 (H_2^\dagger H_1)(H_2^\dagger H_2) + h.c.\right].
\end{eqnarray}

The coefficients in Eq.~\eqref{2HDMV} can be expressed in terms of those in the potential $V(M,H)$, in the notation of \cite{Antola:2009wq}, by:
\bea \label{bTCto2HD}
m_1^2&=&m_M^2\ ,\ m_2^2=\left[m_H^2+\left(2 f^2 c_1+m_M^2 c_3\right) c_3 y_{TC}^2\right] K^2\ ,\ m_{12}^2=\left(f^2 c_1+m_M^2 c_3\right) y_{TC} K\ ,\nonumber\\
\lambda_1&=&\frac{1}{3} \lambda _M\ ,\ \lambda_2=\frac{1}{3}\left(2 c_2 c_3^3 y_{TC}^4+\lambda _H+2 c_3 c_4 y_{TC}^2 \lambda _H+c_3^4 y_{TC}^4 \lambda _M\right) K^4\ ,\nonumber\\ 
\lambda_3&=&\lambda_4=\lambda_5=\frac{1}{6}\left(c_2+c_3 \lambda _M\right) c_3 y_{TC}^2 K^2\ ,\nonumber\\ 
\lambda_6&=&\frac{1}{6}\left(c_2+2 c_3 \lambda _M\right) y_{TC} K\ ,\  \lambda_7=\frac{1}{6}\left[c_4 \lambda _H+c_3^2 y_{TC}^2 \left(3 c_2+2 c_3 \lambda _M\right)\right] y_{TC}K^3\ ,
\eea

The Higgs fields in Eq.~(\ref{2HDMV}) are expressed in terms of the real degrees of freedom by
\be \label{HDs}
H_i = \begin{pmatrix} H_i^+ \\ \frac{1}{\sqrt{2}} (v_i + h_i + i \phi_i) \end{pmatrix} ~,\quad i=1,2\ ; \tan\beta '\equiv \frac{v_1}{v_2}\ .
\ee

The Goldstone boson $G^\pm$ ($G^0$) provides the longitudinal components of the $W^\pm$ ($Z^0$) boson, while $h^0,H^0,A^0$, and $H^\pm$ are the neutral scalars, pseudoscalar, and charged scalar mass eigenstates, respectively: 
\bea
\label{rot2hdm}
& & \begin{pmatrix} h^0 \\ H^0 \end{pmatrix} = \begin{pmatrix} c_{\alpha '} & -s_{\alpha '} \\ s_{\alpha '} & c_{\alpha '} \end{pmatrix} \begin{pmatrix} h_1 \\ h_2 \end{pmatrix} ~, \
\begin{pmatrix} G ^0\\ A^0 \end{pmatrix} = \begin{pmatrix} s_{\beta '} & c_{\beta '} \\ c_{\beta '} & -s_{\beta '} \end{pmatrix}
 \begin{pmatrix} \phi_1 \\ \phi_2 \end{pmatrix} ~,\ \nonumber \\
& & \begin{pmatrix} G^\pm \\ H^\pm \end{pmatrix} = \begin{pmatrix} s_{\beta '} & c_{\beta '} \\ c_{\beta '} & -s_{\beta '} \end{pmatrix} \begin{pmatrix} H_1^\pm \\ H_2^\pm \end{pmatrix} ~.
\eea
The angles $\alpha',\beta'$ differ from $\alpha,\beta$ only because of the extra rotation in Eq.$\eqref{2HDrot}$ that makes only $H_2$ couple to SM fermions.

\end{document}